\documentclass[12pt]{article}
\usepackage{latexsym,amssymb,amsmath}
\begin{document}

\title
{On a construction of self-dual gauge fields in seven dimensions}
\author
{E.K. Loginov\footnote{{\it E-mail address:} ek.loginov@mail.ru}
\medskip\\
\it Department of Physics, Ivanovo State University\\
\it Ermaka St. 39, Ivanovo, 153025, Russia
\bigskip
\and A.N. Grishkov\footnote{{\it E-mail address:} grishkov@ime.usp.br}
\medskip\\
\it IME, University of Sao Paulo\\
CER 005598-300, Sao Paulo, Brazil}
\date{}
\maketitle

\begin{abstract}
We consider gauge fields associated with a semisimple Malcev algebra. We construct a
gauge-invariant Lagrangian and found a solution of modified Yang-Mills equations in seven
dimensions.
\end{abstract}

\section{Introduction}

In Ref.~[1], the 4d (anti-)self-dual Yang-Mills equations were generalized to the
higher-dimensional linear relations
\begin{equation}
c_{mnps}F^{ps}=\lambda F_{mn},
\end{equation}
where the numerical tensor $c_{mnps}$ is completely antisymmetric and $\lambda=const$ is a
non-zero eigenvalue. It is obvious that these equations lead to the full Yang-Mills equation,
via the Bianchi identity. Several anti-self-dual solutions of (1) were found in~[2--6].
Conversely, self-dual solutions of (1) are known. However, if we assume that the tensor
$F_{mn}$ may take values in a Malcev algebra, then such solution may be found. Interestingly
that this solution exactly coincides with the general solution of $7d$ analogue of the $3d$
Euler top equation~[7]. The $7d$ model is regarded as a model describing self-dual membran
instantons~[8].
\par
The paper is organized as follows. Section 2 contains well-known    facts about Cayley-Dickson
algebras, Malcev algebras, and analytic Moufang loops which we use. In Sections 3 gauge fields
associated with a semisimple Malcev algebra are investigated, and a gauge-invariant Lagrangian
of the Yang-Mills type is constructed. In Section 4 solutions of equations of motion in seven
dimensions are found.

\section{Moufang loops and Malcev algebras}

Recall that a loop is a binary system $S$ with an unity element, in which the equations $ax=b$
and $ya=b$ are uniquely solvable for all $a,b\in S$. Moufang loops are distinguished from the
class of all loops by the identity
\begin{eqnarray}
(xy)(zx)=(x(yz))x. \nonumber
\end{eqnarray}
This paper is concerned with connected analytic Moufang loops; that is, connected analytic
manifolds equipped with the Moufang loop structure, in which the binary operations are
analytic. Everywhere below for short we use the term "Moufang loop" in place of the term
"connected analytic Moufang loop".
\par
Moufang loops are closely associated with alternative algebras that are defined by the
identities
\begin{eqnarray}
x^{2}y=x(xy),\qquad yx^{2}=(yx)x. \nonumber
\end{eqnarray}
It is evident from the definition that any associative algebra is alternative. The most
important example of a nonassociative alternative algebra is the real Cayley-Dickson algebra.
Let us recall its construction~(see~[9]).
\par
Let  $A$ be a real linear space equipped with a nondenerate symmetric metric $g$ of signature
$(8,0)$ or $(4,4)$. Choose the basis \{$1,e_{1},...,e_{7}\}$ in $A$ such that
\begin{equation}
g=\text{diag}(1,1,1,1,\pm1,\pm1,\pm1,\pm1),
\end{equation}
and define the multiplication
\begin{equation}
e_{i}e_{j}=-g_{ij}+c_{ij}{}^{k}e_{k},
\end{equation}
where the structure constants $c_{ijk}=c_{ij}{}^{s}g_{sk}$ are completely antisymmetric and
different from 0 only if
\begin{equation}
c_{123}=c_{145}=c_{167}=c_{246}=c_{275}=c_{374}=c_{365}=1. \nonumber
\end{equation}
The multiplication (3) transform $A$ into a linear algebra that is called the Cayley-Dickson
algebra.
\par
In general, the commutator algebra $A^{(-)}$ of the alternative algebra $A$ is not a Lie
algebra. Instead of the Jacobi identity, $A^{(-)}$ satisfies the identity
\begin{eqnarray}
J(x,y,[x,z])=[J(x,y,z),x],
\end{eqnarray}
where $J(x,y,z)$ is so-called Jacobian of the elements $x,y,z$
\begin{eqnarray}
J(x,y,z)=[[x,y],z]+[[y,z],x]+[[z,x],y].
\end{eqnarray}
An anticommutative algebra whose multiplication satisfies the identity (4) is called a Malcev
algebra. Everywhere below we use the term "Malcev algebra" in place of the term
"finite-dimensional real Malcev algebra".
\par
Malcev algebras and alternative algebras are closely associated~[10--12]. Any non-Lie simple
Malcev algebra is isomorphic to the quotient algebra $A^{(-)}/\mathbb{R}$, where $A$ is a
Cayley-Dickson algebra. Any semisimple Malcev algebra $M$ is decomposed in the direct sum
$M=N(M)\oplus J(M)$ of Lie center $N(M)$ and ideal $J(M)$ generated by Jacobians. Besides,
$N(M)$  is a semisimple Lie algebra and $J(M)$ is a direct sum of non-Lie simple Malcev
algebras. In particular, it follows from here that any semisimple Malcev algebra is embedded
in the commutator algebra of an appropriate alternative algebra.
\par
Further, recall that a derivation of a Malcev algebra $M$ is a linear transfor\-mation $D$ of
$M$ satisfying
\begin{eqnarray}
D[x,y]=[Dx,y]+[x,Dy] \nonumber
\end{eqnarray}
for all $x,y\in M$. We may write it in an equivalent form
\begin{eqnarray}
[D,T_x]=T_{Dx},
\end{eqnarray}
where $T:M\to End(M)$ is the regular representation of $M$, such that $T_xy=[x,y]$.
\par
We denote by $Lie(M)$ the Lie algebra generated by all operators $T_x$. If $M$ is a semisimple
Malcev algebra, then $Lie(M)$ is a direct sum of the derivation subalgebra
$Der(M)=Der(N)\oplus Der(J)$ and vector subspace $T(M)\simeq J(M)$. In addition, the
subalgebra $Der(N)\simeq N(M)$ and the subalgebra $Der(J)$ is linearly generated by the
operators
\begin{eqnarray}
D(x,y)=T_{[x,y]}+[T_x,T_y],
\end{eqnarray}
where $x,y\in J(M)$. With the regular representation $T$ we may connect the bilinear form
$(x,y)=\text{tr}(T_{x}T_{y})$. It is clear that the form $(x,y)$ is symmetric $(x,y)=(y,x)$.
In addition, it follows from (4) that
\begin{eqnarray}
([x,y],z)=(x,[y,z]),
\end{eqnarray}
for any $x,y,z\in M$. The form $(x,y)$ is called a Killing form.
\par
The regular representation of Malcev algebras can be extended. Let $M$ be a Malcev algebra,
$V$ be a real vector space, and $\tau:M\to End(V)$,  $x\to\tau_x$ a linear mapping. Then
$\tau$ is called a representation of $M$ if the algebra defined on the space $M\oplus V$ by
\begin{eqnarray}
[x+v,y+w]=[x,y]+\tau_x w-\tau_y v \nonumber
\end{eqnarray}
is a Malcev algebra. In this case $V$ is called a Malcev module. It is  known~[13--14] that
every representation of a semisimple Malcev algebra is completely reducible. Any irreducible
Malcev module is either Lie or the regular bimodule for a nonassociative simple Malcev algebra
or $sl(2)$-module of dimension 2 such that $\tau_x=\bar x$, where $\bar x$ is the adjoint
matrix to $x\in sl(2)$.
\par
Finally, we note that there exists a correspondence between Moufang loops and Malcev algebras,
which generalizes the classical Lie correspondence between Lie groups and Lie
algebras~[15--17]. Namely, there exists an unique, to within isomorphism, simply connected
Moufang loop $S$ with a given tangent Malcev algebra, and any Moufang loop with the same
tangent algebra is isomorphic to the quotient algebra $S/N$ where $N$ is a discrete central
normal subgroup of $S$. Any simply connected semisimple Moufang loop is decomposed in the
direct product of a semisimple Lie group and a simple nonassociative Moufang loops each of
which is analytically isomorphic to one of the spaces ${\mathbb S}^7$, ${\mathbb
S}^3\times{\mathbb R}^4$, or ${\mathbb S}^7\times{\mathbb R}^7$. Actually, any simply
connected simple nonassociative Moufang loop is isomorphic to the loop of elements of norm 1
in the Cayley-Dickson algebra over ${\mathbb R}$ or ${\mathbb C}$.
\par
Using the previously mentioned correspondence between Moufang loops and Malcev algebras, we
can define a notion of the representation of Moufang loop. Let $S$ be a Moufang loop and $M$
be its tangent Malcev algebra. The representation $M\to End(V)$ of $M$ induces the mapping
$\widetilde S\to Aut(V)$ of the simply connected Moufang loop $\widetilde S$, locally
isomorphic to $S$, into the simply connected Lie group $\widetilde G\subseteq Aut(V)$. This
mapping can extend into the group $G\simeq\widetilde G/\widetilde G_0$, where the subgroup
$\widetilde G_0$ lies in the discrete center of $\widetilde G$. Therefore if
$S\simeq\widetilde S/\widetilde S_0$ and $\widetilde S_0$ is isomorphic to a subgroup of
$\widetilde G_0$, then there exists the mapping $S\to G$ induced by the representation of $M$.
We shall call this mapping by a representation of $S$. In particular, if $S$ is a
nonassociative compact simple Moufang loop, then $\widetilde G\simeq Spin(7)$ and $G\simeq
SO(7)$.

\section{Nonassociative gauge fields}

Suppose $S$ is a semisimple Moufang loop, $M$ is its tangent Malcev algebra, and $A_m(x)$ is a
vector field taking values in $M$ and defined in an Euclidean or pseudo-Euclidean space.
Further, let
\begin{eqnarray}
S\to G\subseteq Aut(V)
\end{eqnarray}
be a representation of $S$, and $H$ be a subgroup of $G$ locally isomorphic to the group
$Int(S)$ of inner automorphisms of $S$. Define the field $\psi(x)$ taking values in $V$ and
its covariant derivative
\begin{eqnarray}
\hat D_m\psi=\left(\partial_m+\hat A_m\right)\psi, \nonumber
\end{eqnarray}
where $\hat A_m\in End(V)$ is an image of $A_m$ corresponding to the representation (9). As
usually, we require that $\hat D_m\psi$ has transformation properties the same as the field
$\psi$, i.e.
\begin{align}
\psi&\to\psi'=\hat u\psi,
\nonumber\\
\hat D_m\psi&\to\hat D_m'\psi'=\hat u\hat D_m\psi,
\end{align}
where $\hat u(x)$ is function taking values in the subgroup $H$.
\par
Consider an infinitesimal gauge transformation $\hat u=1+\hat\varepsilon$. It is obvious that
$\hat\varepsilon\in Der(M)$. If the field $A_m$ takes values in the Lie center $N(M)$, then
the condition (10) defines the usual transformation rule of gauge fields
\begin{eqnarray}
A_m\to A_m+[\varepsilon, A_m]-\partial_m\varepsilon,
\end{eqnarray}
where $\varepsilon$ is an isomorphic prototype of $\hat\varepsilon$ in $N(M)$. If $A_m$ is an
element of the ideal $J(M)$ generated by Jacobians, then the situation is different. In this
case the operator $\hat\varepsilon$ has not a prototype in $M$. On the other hand, the
representation of $J(M)$ induced by the representation of $S$ is regular. Therefore we may use
the identity (6) and get the following transformation rule
\begin{eqnarray}
\left.\aligned
\partial_m&\to\partial_m-\partial_m\hat\varepsilon,\\
A_m&\to A_m+\hat\varepsilon A_m\\
\endaligned\right\}.
\end{eqnarray}
Obviously, the transformation (12) of $A_m$ induces a transformation of $\hat A_m$ according
to the rule
\begin{eqnarray}
\hat A_m\to\hat u\hat A_m\hat u^{-1}.
\end{eqnarray}
\par
Now we go on to a construction of the gauge-invariant Lagrangian. To this end  define tensor
fields in $M$ such that their images in $End(V)$ are transformed by the adjoined
representation of $G$. This request is satisfied if we suppose
\begin{align}
F_{mn}&=\partial_mA_n-\partial_nA_m+[A_m,A_n],
\nonumber\\
J_m&=\frac{1}{12}c_{mnps}J(A^n,A^p,A^s), \nonumber
\end{align}
where $c_{mnps}$ is a completely antisymmetric numerical tensor. Indeed, it can easily be
checked that the infinitesimal transformations (11) and (12) induce the transformations
\begin{alignat}{2}
F_{mn}&\to F_{mn}+\hat\varepsilon F_{mn},&\qquad J_{m}&\to J_{m}+\hat\varepsilon J_{m},
\nonumber\\
\hat F_{mn}&\to \hat u\hat F_{mn}\hat u^{-1},&\qquad \hat J_{m}&\to \hat u\hat J_{m}\hat
u^{-1},
\end{alignat}
where $\hat F_{mn}$ and $\hat J_m$ are images of $F_{mn}$ and $J_m$, which are defined by the
representation (9). Note that the tensor $F_{mn}$ extend a notion of the Yang-Mills field
strength. Conversely, the field $J_m$ has not a prototype in the Yang-Mills theory. The field
$J_m$ is not zero only if $M$ is a non-Lie Malcev algebra.
\par
We define the Lagrangian
\begin{eqnarray}
L=\frac1{8g^2}\text{tr}(\hat F_{mn}\hat F^{mn}+\hat J_s\hat A^s).
\end{eqnarray}
If $M$ is a Lie algebra, then the Lagrangian (15) coincides with the Lagrangian of Yang-Mills
and hence it is gauge-invariant. On the other hand, every representation of a semisimple
Malcev algebra is completely reducible. There\-fore it is enough to prove $H$-invariance of
the Lagrangian (15) only if $M$ is a non-Lie simple Malcev algebra. But in this case the
gauge-invariance of the Lagrangian (15) follows from (13) and (14).

\section{Solutions of the equations of motion}

Let $M$ be a non-Lie simple Malcev algebra. Then $M\simeq A^{(-)}/\mathbb R$, where $A$ is a
Cayley-Dickson algebra, and hence the algebra $M$ admits the basis $e_{1},\dots,e_{7}$ with
the multiplication table
\begin{eqnarray}
[e_{i},e_{j}]=2c_{ij}{}^{k}e_{k}. \nonumber
\end{eqnarray}
In this basis the operators $T_{e_i}$ are represented in the form
\begin{eqnarray}
T_{e_i}=c_{i}{}^{jk}e_{jk},
\end{eqnarray}
where $e_{jk}$ are generators of the Lie algebra $Lie(M)\simeq so(7)$ or $so(3,4)$ with the
matrix elements
\begin{eqnarray}
(e_{jk})^a_b=g_{jb}\delta^a_k-g_{kb}\delta^a_j.
\end{eqnarray}
Using the representation (16) and the identity (17), we can find the Killing form
\begin{eqnarray}
(e_i,e_j)=-24g_{ij}.
\end{eqnarray}
Substituting (18) in the Lagrangian (15), we easily get the equations of motion
\begin{eqnarray}
\partial^mF_{mn}+[A^m,F_{mn}]=J_n.
\end{eqnarray}
Using the definition (5), we prove that every solution of the equations
\begin{eqnarray}
F_{mn}=\frac14c_{mnps}F^{ps}
\end{eqnarray}
is a solution of the equations of motion (19).
\par
Further, let $H$ be a Lie group locally isomorphic to the group $Int(M)$ of inner
automorphisms of $M$. Then $H\simeq G_2$ (or $H\simeq G'_2$ in noncompact case). Define by
\begin{eqnarray}
J(e_i,e_j,e_k)=12c_{ijk}{}^le_l \nonumber
\end{eqnarray}
the completely antisymmetric $H$-invariant tensor $c_{ijkl}$. If we write out its nonzero
components
\begin{eqnarray}
c_{4567}=c_{2367}=c_{2345}=c_{1357}=c_{1364}=c_{1265}=c_{1274}=1, \nonumber
\end{eqnarray}
then it is easy to prove that the tensors $c_{ijk}$ and $c_{ijkl}$ satisfy the following
identities
\begin{align}
c_{mni}c_{ps}{}^{i}&=g_{mp}g_{ns}-g_{ms}g_{np}+c_{mnps},\\
c_{mni}c_{psj}{}^{i}&=c_{ps[m}g_{n]j}+c_{sj[m}g_{n]p}+c_{jp[m}g_{n]s},\\
c_{mnij}c_{p}{}^{ij}&=4c_{mnp},\\
c_{mnij}c_{ps}{}^{ij}&=4g_{mp}g_{ns}-4g_{ms}g_{np}+2c_{mnps}.
\end{align}
\par
Note that if the gauge fields are defined in an Euclidean space and $M$ is a compact algebra,
then any solution of (20) correspond to a minimum of the functional of action. Indeed, using
the identities (8) and (24), we can shown that
\begin{eqnarray}
L=\frac1{8g^2}\text{tr}\left[\frac23\left(\hat F_{mn}-\frac14c_{mnps}\hat
F^{ps}\right)^2+\partial^m\hat I_m\right], \nonumber
\end{eqnarray}
where the prototype of $\hat I_m$ in $M$ has the form
\begin{eqnarray}
I_m=c_{mnps}\left(A^n\partial^pA^s+\frac23(A^nA^p)A^s\right). \nonumber
\end{eqnarray}
Since the variation of field on the boundary of volume vanish in the deduction of equations of
motion, we get the required statement.
\par
Now we turn to a search of solutions of the equations (20). Suppose that the field $F_{mn}$ is
defined in the space $M$ with the metric (2). We choose the ansatz
\begin{eqnarray}
A_m(x)=\frac{c_{mij}x^j}{\lambda^2+x^2}e^i,
\end{eqnarray}
where $x^2=x_kx^k$. Using the identities (21)--(22), we get the following expression of field
strength
\begin{eqnarray}
F_{mn}(x)=\frac{2c_{mni}(\delta^i_j\lambda^2+2x^ix_j)}{(\lambda^2+x^2)^2}e^j.
\end{eqnarray}
By the identity (23), it follows that the tensor (26) is self-dual
\par
The obtained solution may be generalized. We choose the ansatz
\begin{eqnarray}
A_m(x)=6c_{mij}x^iB^j(u),
\end{eqnarray}
where depending on $u=\lambda^2+x^2$ the vector $B^j$ takes values in $M$. If the vector $B_i$
satisfies the identity
\begin{eqnarray}
\frac{dB_i}{du}=\frac12c_{ijk}[B^j,B^k],
\end{eqnarray}
then the field strength again is self-dual
\begin{eqnarray}
F_{mn}(x)=12c_{mnj}\left(B^j+\frac{dB^j}{du}x^2+\frac{dB_i}{du}x^ix^j\right). \nonumber
\end{eqnarray}
A particular solution of the equation (28) can find if to compare the right parts of (25) and
(27). General solution of (28) was found in the work~[7].

\bigskip\medskip\par\noindent
{\bf Acknowledgements}
\medskip\par\noindent
E.K. Loginov is grateful to the foundations RFBR 06-02-16140 and FAPESP 06/61138-6 for
support. A.N. Grishkov thanks FAPESP 05/60337-2 and 05/00666-2 for partial support.

\end{document}